\documentclass[12pt,reqno]{article}
\usepackage{amsmath,hyperref,amssymb}
\usepackage{subfigure,graphicx,url}
\usepackage{cite}

\newcommand{\bea}{\begin{eqnarray}}
\newcommand{\eea}{\end{eqnarray}}
\newcommand{\be}{\begin{equation}}
\newcommand{\ee}{\end{equation}}
\newcommand{\eeq}{\end{equation}}
\newcommand{\beq}{\begin{equation}}

\allowdisplaybreaks \numberwithin{equation}{section}

\DeclareSymbolFont{AMSa}{U}{msa}{m}{n}
\DeclareSymbolFont{AMSb}{U}{msb}{m}{n}
\DeclareMathSymbol{\fieldR}{\mathalpha}{AMSb}{"52}

\def\beq{\begin{equation}}
\def\eeq{\end{equation}}
\def\be{\begin{equation}}
\def\ee{\end{equation}}
\def\bea{\begin{eqnarray}}
\def\eea{\end{eqnarray}}

\begin{document}
\begin{flushright} \small
\end{flushright}
\bigskip
\begin{center}
 {\large\bfseries  Vortex Lattices and Crystalline Geometries  }\\[3mm]

 \

Ning Bao$^*$, Sarah Harrison$^*$, Shamit Kachru$^{*}$ and Subir Sachdev$^{\dag}$   \\[5mm]
 
 {\small\slshape
 * SITP, Department of Physics\\
 and\\
 Theory Group, SLAC \\
 Stanford University\\
 Stanford, CA 94305, USA \\
\medskip
 \dag Department of Physics\\
  Harvard University\\
 Cambridge, MA 02138, USA\\
\medskip
 {\upshape\ttfamily ningbao@stanford.edu, sarharr@stanford.edu, skachru@stanford.edu, sachdev@g.harvard.edu}
\\[3mm]}
\end{center}

\

\

\vspace{1mm}  \centerline{\bfseries Abstract}
\medskip

We consider $AdS_2 \times R^2$ solutions supported by a magnetic field, such as those which arise in the near-horizon limit of magnetically charged $AdS_4$ Reissner-Nordstrom black branes.  In the presence of an electrically charged scalar field, such magnetic solutions can be unstable to spontaneous formation of a vortex lattice.  We solve the coupled partial differential equations which govern the charged scalar, gauge field, and metric degrees of freedom to lowest non-trivial order in an expansion around the critical point, and discuss the corrections to the free energy and thermodynamic
functions arising from the formation of the lattice.  We describe how such solutions can also be
interpreted, via S-duality, as characterizing infrared crystalline phases of conformal field theories doped by a chemical potential,
but in zero magnetic field; the doped conformal field theories are dual to geometries that
exhibit dynamical scaling and hyperscaling violation.

\bigskip
\newpage

\tableofcontents

\newpage


\section{Introduction}

A topic of recent interest has been the holographic description of phases of quantum field theory with spatial anisotropy and/or inhomogeneity \cite{DH,branelattices,Maeda,ooguri,Tong,Benincasa,gauntlettspace,Bianchi,Nori,Horowitz,Erdmenger,Rozali,DonosHartnoll,Cremonini,Vegh}.  This is motivated in part by the crucial role that momentum relaxation due to inhomogeneities plays in transport phenomena in condensed matter systems, and in part by intrinsic interest in the rich physics of such phases.  

Our goal in this work is twofold.  On the one hand, as an extension of the ideas discussed in \cite{Sachdevnew}, we would like to illustrate the emergence of crystalline ground states (`solids') 
in conformal field theories doped by a chemical potential coupling to a globally conserved U(1) charge, but in zero magnetic field.
In 2+1 dimensions, monopole operators associated with the global U(1) symmetry \cite{Willett,Sachdevnew,Pufu}
serve as order
parameters for solid phases in doped CFTs.  Electric-magnetic duality allows one to find a dual
description where the magnetic degrees of freedom are manifested in terms of electrically
charged operators.  In the bulk gravitational description, this allows us to view the formation
of the solid by studying vortex lattice formation in the theory of a charged scalar moving in a
background magnetic field. A crucial advantage of studying the solid phases of doped CFTs by using this dual charged scalar is
that the Dirac quantization condition on the monopole charge 
translates into an exact commensurability relation between the area of the unit cell 
of the crystal and the density of doped charges \cite{Iqbal,Sachdevnew}.

On the other hand, an open problem in the study of holographic lattices has been to find,
analytically, gravitationally back-reacted solutions for a crystalline lattice of dimension $d > 1$.  
This has largely been because of the relative difficulty of solving coupled systems of partial differential
equations, instead of the ordinary differential equations which normally govern simple backgrounds in gauge/gravity duality.
Here, we give an example of such a crystalline metric in $d=2$.  Our work builds on the earlier papers
\cite{Maeda}, which found an elegant solution for a vortex lattice in the probe approximation, and
\cite{Erdmenger}, where the backreaction of such a lattice on bulk gauge fields was studied in a different setting.  

The organization of this paper is as follows.  In \S2, we review the basic unperturbed $AdS_2 \times R^2$ solution.  In \S3, we incorporate a charged scalar field and describe the vortex lattice solution.  In \S4, we describe the basic physics visible in the perturbative vortex lattice solution.
In \S5, we characterize how such a lattice could also emerge in the IR geometry of a gravitational solution which exhibits dynamical scaling with hyperscaling violation, as the S-dual of a doped CFT in zero magnetic field.  Possible directions for future research
are discussed in \S6.

\section{Magnetic $AdS_2 \times R^2$ solutions}

Consider the theory with action
\begin{equation}
S = \int ~d^4x~\sqrt{-g} \left( R - {1\over 4} F_{\mu\nu}F^{\mu\nu} - 2\Lambda \right)~
\end{equation}
where $\Lambda < 0$.
It has $AdS_2 \times R^2$ solutions with metric
\begin{equation}
\label{AdS2}
ds^2 = L^2 \left(-{dt^2 \over r^2} + {dr^2 \over r^2}  + dx^2 + dy^2 \right) ~
\end{equation}
where 
\begin{equation}
{1\over L^2} = - 2\Lambda~.
\end{equation}

The gauge field supporting the solution is
\begin{equation}
F_{xy} = Q_{m} dx \wedge dy
\end{equation}
with $Q_m$ fixed in terms of the AdS radius by the equation
\begin{equation}
Q_m = \sqrt{2} L~.
\end{equation}
In particular, this means that for these solutions, fixing the magnetic field fixes also the cosmological constant and the AdS radius.

In addition to its intrinsic interest, this solution arises as the near-horizon geometry of extremal magnetically charged AdS/Reissner-Nordstrom
black branes with $AdS_4$ asymptotics.  In this context, the $AdS_2$ near-horizon region has played a crucial role in elucidating the 
non-Fermi liquid behavior of probe fermions \cite{SS,MIT,Leiden} scattering off the bath of  locally critical excitations represented by the AdS$_2$ geometry \cite{Faulkner,Subir}.

\section{The vortex lattice}

Our interest is not in the pure $AdS_2$ solution (\ref{AdS2}).  
We wish to include also an electrically charged scalar field, $\psi$, in the full action.  In part, this is because a generic such theory
could include such scalars; in part, it is motivated by the duality considerations to be described in \S5.

In any case, here, we will see that in some ranges of parameters, the charged scalar will qualitatively change the
IR physics. The simplest case in which we can see this effect will be directly in the $AdS_2\times R^2$ background of \S ~2. We will impose a hard wall cutoff at $r=r_0$ in the deep IR, along with suitable boundary conditions, to be described below.  We can think of $r_0$ as a proxy for a `confinement scale' or a `temperature.'  Tuning the magnetic field relative to the `temperature' will trigger the scalar instability.

After including the $\psi$ coupling to the gauge field the action becomes
\be
S=\int d^4x\sqrt{-g}\left(R-2\Lambda-{1\over 4}F^2-|\nabla_\mu \psi|^2-m^2|\psi|^2-\lambda |\psi|^4\right )~.
\ee
We have defined $\nabla_\mu=\partial_\mu+ieA_\mu$, where $e$ is the electric charge of the scalar field. From this point on we will set $e=1$.
This action has a stress-energy tensor of the form
\be
T_{\mu\nu}=-{g_{\mu\nu}\over 2}\mathcal{L}_{mat}+{1\over 2}F_{\mu\lambda}F_\nu^\lambda+e^2A_\mu A_\nu|\psi|^2+{1\over 2}[\partial_\mu\psi\partial_\nu\psi^*+ie\psi(A_\mu\partial_\nu+A_\nu\partial_\mu)\psi^*+h.c.]
\ee
where
\be
\mathcal{L}_{mat}={1\over 4}F^2+|\nabla_\mu \psi|^2+m^2|\psi|^2+\lambda |\psi|^4.
\ee

We may  expand the magnitude of $\left|\nabla_\mu \psi \right|^2$ as
\beq
\left|\nabla_\mu \psi \right|^2=\left|\partial_\mu\psi\right|^2+iA_\mu(\psi\partial^\mu\psi^* -\psi^*\partial^\mu\psi)+A^2\left|\psi\right|^2.
\eeq
At this point, we can calculate the Euler-Lagrange equation for $\psi$ by differentiating with respect to $\psi^*$:
\beq\label{eq:psi_eom}
\partial_\mu(\sqrt{-g}\nabla^\mu\psi)=-\sqrt{-g}(iA^\mu\nabla_\mu\psi-m^2\psi-2\lambda|\psi|^2\psi),
\eeq
and the equation of motion for the gauge field,
\be\label{eq:A_eom}
{1\over\sqrt{-g}}\partial_\mu(\sqrt{-g}F^{\mu\nu})=i(\psi\partial^\nu\psi^*-\psi^*\partial^\nu\psi)+2A^\nu|\psi|^2.
\ee

In addition to these equations of motion, we will also need to solve the Einstein equations,
\be
R_{\mu\nu}-\frac{(R-2\Lambda)}{2}g_{\mu\nu}=T_{\mu\nu},
\ee
when we include backreaction of the $\psi$ condensate on the gauge field and the metric.

We will expand perturbatively in a small parameter $\epsilon$ around the solution $\psi=0$ with background gauge field of the form
\begin{equation}
\label{gaugechoice}
A_x = Q_c  y, A_y = 0.
\end{equation}
The scalar field in the competing vortex phase  will itself be of order $\epsilon$.
For fixed $r_0$ and boundary conditions (to be discussed below), we will choose $Q_c$ to be just at the onset for the transition to forming vortices. At this critical value of the magnetic field, the $\psi=0$ solution will be degenerate with a vortex lattice solution. As we increase the magnetic field to slightly above its critical value, $\psi=0$ will no longer be the preferred solution, and the vortex lattice will be preferred.  As is familiar, the onset of the transition is signalled by the existence of a purely normalizable solution for 
$\psi$ that respects the IR boundary conditions.


We can parametrize the backreaction of the scalar on the gauge sector through a perturbative expansion in the distance away from the critical field.
The scalar will have the form
\be\label{eq:psi_expansion}
\psi(r,x,y)=\epsilon\psi_1(r,x,y)+\epsilon^3\psi_3(r,x,y)+\ldots
\ee
and the gauge field will have the form
\bea\label{eq:A_expansion}
A_x(r,x,y)&=&Qy+\epsilon^2a^x_2(r,x,y)+\ldots,\\
A_y(r,x,y)&=&\epsilon^2a^y_2(r,x,y)+\ldots,
\eea
with $A_t=A_r=0$. When we consider lattice solutions which are periodic in $x$ and $y$, the backreaction of $\psi$ on the gauge field will require both $A_x$ and $A_y$ to be nonzero at $O(\epsilon^2)$, with both $x$ and $y$ dependence. 

A similar statement holds for the metric at $O(\epsilon^2)$.
Our metric ansatz, to $O(\epsilon^2)$, will be
\be
ds^2=L^2\left \{{1\over r^2}((-1+\epsilon^2 a(r,x,y))dt^2+(1+\epsilon^2 a(r,x,y))dr^2)+(1+\epsilon^2 b(r,x,y))(dx^2+dy^2)\right \}.
\ee
Because at zeroth order in epsilon the $AdS_2\times R^2$ metric is exactly supported by the magnetic field (i.e. the gauge field is not a probe), we find it necessary to include metric backreaction once we backreact on the gauge field.  This distinguishes our situation from that considered in e.g. \cite{Erdmenger}.

The radial magnetic field will be 
\be
B_r=Q+\epsilon^2(\partial_ya^x_2(r,x,y)-\partial_xa^y_2(r,x,y))
\ee

In general, when we backreact on the magnetic field, we may expect there to be a non-normalizable piece at order $\epsilon^2$, i.e. $A_x(r\to 0)= (Q+\delta Q\epsilon^2)y$. This shifts the naive critical value of the field at the transition. However, because the critical point is actually only dependent on the dimensionless combination $Q/r_0^2$, we can (and will) impose that there is no non-normalizable correction to the gauge field in our backreacted solutions. That is, we will set $\delta Q=0$. The value of the critical point will still have an $O(\epsilon^2)$ shift; it will manifest itself as an $O(\epsilon^2)$ shift in the location of the hard wall, $r_0\to r_0+\delta r_0\epsilon^2$. These two scenarios are equivalent; in both cases we should think of the backreaction of $\psi$ on the metric and gauge field as inducing a shift in the dimensionless parameter which controls the critical point at $O(\epsilon^2)$.

\subsection{Basic droplet}
We now examine the solutions of the field equation for $\psi$, in the limit where we can neglect the back-reaction of $\psi$
on the gauge field and on the metric.  (This will be at order $\epsilon$.) Very similar equations have been examined in the literature on vortices in holographic
superconductors \cite{Johnson, HHH, Maeda}. 
The basic building block for the solutions we will study is the ``droplet" solution of \cite{Johnson}.

We will begin by setting $\lambda=0$ in the potential for the scalar and proceed with the metric and the mass of the (dualized) monopole field unspecified.  Both of these will affect the radial solution for the scalar, but we will see that the spatial part of $\psi_1$ decouples from the radial equation for all metrics we might consider, and so we can find the basic form of the droplet solution while leaving the metric general.

For metrics with components which only depend on $r$ and for which $g^{xx}=g^{yy}$, we can solve this equation by separation of variables, assuming that
\beq
\psi_1=\rho_0\rho(r)g(y)e^{ikx}~,
\eeq
where $\rho_0$ is an overall constant. Inserting our choice of gauge (\ref{gaugechoice})
yields, after some algebraic manipulation,
\bea
\frac{1}{\rho_n(r)}\left (\frac{g^{rr}}{g^{xx}}\rho_n''(r)+\frac{1}{\sqrt{-g}g^{xx}}\frac{\partial}{\partial r}(\sqrt{-g}g^{rr})\rho_n'(r)\right )-\frac{m^2}{g^{xx}} \\\nonumber
= -\frac{1}{g_n(y)}\left(g_n''(y)-(Qy+k)^2g_n(y)\right )=-\lambda_n,
\eea
where $\lambda_n$ is the eigenvalue from the separation of variables.
First we will consider the equation for $g(y)$, which will yield the basic droplet solution. This solution will only exist in the parameter ranges which admit a normalizable solution to the radial equation; we will discuss this in the next section. The equation for $g$ becomes,
\beq
\label{SHO}
g''_n-(Qy+k)^2g_n=\lambda_ng_n.
\eeq

Now, redefining $Y=\sqrt{Q}(y+\frac{k}{Q})$, the $g_n$ equation becomes
\beq
g''_n(Y)-\left(Y^2+\frac{\lambda_n}{Q}\right )g_n(Y)=0.
\eeq
Solving, we get that
\beq
g_n(Y)=c_+ D_{\nu_+}(\sqrt{2}Y)+c_-D_{\nu_-}(i\sqrt{2}Y)
\eeq
where $c_\pm$ are constants, $\nu_\pm=\frac{1}{2}\left (-1\pm\frac{\lambda_n}{Q}\right )$, and $D_\nu(x)$ is the parabolic cylinder function.

The reader may recognize the differential equation for $g_n$, (\ref{SHO}), as the same eigenvalue problem that arises in 
the study of the quantum mechanics of the simple harmonic oscillator.  More properly, this is the case for appropriate choices
of the separation constant.  In these cases, we can write the (normalizable) solution for $g_n$ in terms of the familiar
Hermite polynomials:
\beq
g_n=e^{-Y^2/2}H_n(Y),
\eeq
with eigenvalues $\lambda_n=2Q(n+1/2)$.  
The nth eigenvalue here characterizes the nth Landau level of the $\psi$ particles.
The ``droplet" solutions with this shape were first discussed in the series of papers \cite{Johnson}, in a related but distinct context. The single droplet solution is when $n=0$, which is just a Gaussian  centered at $y=-k/Q$, $g(y)=e^{-Y^2/2}$. Note that $g_n=$ constant is not a solution to the equations of motion.

\subsection{Vortex lattice}

Of more interest to us is a solution which preserves some discrete subgroup of the translation invariance of the original system.
The basic droplet of \S3.1 breaks translations entirely.  However, more symmetric solutions can be obtained by taking linear combinations
of droplets, which still solve the (linearized) equations of motion neglecting back reaction.  

A vortex lattice can be constructed as follows \cite{Maeda}, using the zeroth Landau level solutions for the $\psi$ field.  The basic solution is
\begin{equation}
\psi_0(y;k) = e^{-{Y^2\over 2}} = e^{-{Q\over 2} (y + {k \over Q})^2}~.
\end{equation}
An appropriate superposition to give a lattice in the $x-y$ plane is
\begin{equation}
\label{latticesol}
\Psi_{\rm lat}(x,y) =
{1 \over L}  \sum_{l=-\infty}^{\infty} c_l e^{i k_l x} \psi_0(y;k_l)
\end{equation}
where
\begin{equation}
c_l \equiv e^{-i \pi {v_2 \over v_1^2} l^2},~~k_l \equiv {2\pi l \over v_1}\sqrt{Q}
\end{equation}
for ${\it arbitrary}$ $v_1$ and $v_2$.
 
 One can write this in terms of the elliptic theta function $\theta_3$:
 \begin{equation}
 \theta_3(v,\tau) \equiv  \sum_{l=-\infty}^{\infty} q^{l^2} z^{2l}, ~~q \equiv e^{i\pi\tau},~z\equiv e^{i\pi v}
 \end{equation}
 as
 \begin{equation}
 \psi_1(x,y,r) = \rho_0\rho(r) \Psi_{lat}(x,y),~~
 \Psi_{lat}(x,y) \equiv e^{-Q y^2 \over 2} \theta_3(v,\tau)
\end{equation}
with
\begin{equation}
v \equiv {\sqrt{Q} (x+iy) \over v_1},~~\tau \equiv {{2\pi i - v_2} \over v_1^2}~.
\end{equation}

That the solution (\ref{latticesol}) represents a lattice is now evident from the basic properties
of the elliptic theta function.  For instance
\begin{equation}
\theta_3 (v+1, \tau) = \theta_3 (v,\tau)
\end{equation}
and
\begin{equation}
\theta_3(v+\tau, \tau) = e^{-2\pi i(v + \tau/2)} \theta_3 (v,\tau)
\end{equation}
implying that
$\Psi_{\rm lattice}$ returns to its value (up to a phase) upon translation by the lattice generators
\begin{equation}
{\bf a} = {1\over \sqrt{Q}} v_1 \partial_x, ~~{\bf b} = {1\over \sqrt{Q}}\left({2\pi \over v_1} \partial_y + {v_2 \over v_1} \partial_x\right)~.
\end{equation}
These have been fixed such that the area of a unit cell is $2\pi/Q$, containing exactly one flux quantum.
It is this quantization condition which translates, in the electromagnetic dual, to the commensurability condition between
the area of the unit cell and the density of doped charges \cite{Iqbal,Sachdevnew}.

That $\Psi_{lat}$ should be called a vortex lattice, despite the fact that it is composed of an array of the droplet solutions
of \cite{Johnson}, is evident from the fact that $\theta_3$ vanishes on the lattice spanned by 
half-integral multiples of the lattice generators
(giving rise to vortex cores), and has a phase rotation of $2\pi$ around each such zero.

Some common lattice shapes are obtained by choosing particular values of the parameters $v_1,v_2$. A rectangular lattice can be obtained by setting $v_2=0$. In this case all coefficients in equation (\ref{latticesol}) are equal, $c_l=c=1$. The ratio of length to width of the rectangle is parametrized by $v_1$. For the special choice $v_1=\sqrt{2\pi}$, the lattice is square. Another special choice is $v_2={1\over 2}v_1^2$; this yields a rhombic lattice. In this case $c_l=1$ for $l\equiv0\bmod 2$ and $c_l=-i$ for $l\equiv 1\bmod 2$. For the special case $v_1=2\sqrt{\pi}$ the rhombus is square (but now rotated $45^\circ$ w.r.t the $x$ axis), and for $v_1={2\sqrt{\pi}\over 3^{1\over 4}}$ the lattice is composed of equilateral triangles (though the unit cell is still a rhombus).

At this point, nothing has fixed the ``moduli" $v_1, v_2$ of the vortex lattice, nor the overall magnitude $\rho_0$ of $\psi_1$.  In standard Landau-Ginzburg theories, apparently the triangular lattice is preferred.
One could find preferred shapes in the approach here by including leading non-linearities in $|\psi|$ in the free energy, and minimizing the free energy density.  It might be interesting to do this, while introducing parameters to vary that could lead to phase transitions in the preferred lattice shape.

\subsection{The radial equation and boundary conditions}

Now we consider the differential equation for $\rho(r)$. At order $\epsilon$ we are still in an $AdS_2\times R^2$ background, which means that $\psi$ should scale as a power law in $r$. Choosing the solution that vanishes at the boundary, we get
\be
\rho(r)=r^\alpha
\ee
where $\alpha={1\over 2}\left(1+\sqrt{1+4(Q+m^2L^2)}\right)$. 

At the hard wall cutoff, $r=r_0$, we will need to impose a consistent set of boundary conditions.  One way to do this is to consider a method very similar to the prescription of \cite{RS}. We will add a mirror image of the spacetime to the other side of the wall and glue them together at the IR boundary, $r=r_0$. Thus, we have two asymptotic UV boundaries (in our coordinates at $r=0$ and $r=2r_0$) and mirror solutions for the metric and the fields on either side of the wall. We will require the metric and fields to be continuous at the wall, but their derivatives will have a discontinuity.  That is, we impose Israel junction conditions at the wall, including any localized energy-momentum sources present there.
At the end of the day, we can quotient by the $Z_2$ symmetry to leave just one copy of the desired space-time.

 In order to support the discontinuity at the IR wall and thus solve the equations of motion at the wall, there must be a source of stress-energy at $r=r_0$. Therefore we will add an action, $S_{wall}$, localized to the wall and solve the equations of motion. One way that this is different from the situation discussed in \cite{RS} is that while those authors needed only to add a localized cosmological constant to the wall (as everything was only a function of the radial variable), we now have spatial $(x,y)$ dependence, so our boundary action must also have spatial dependence, $S_{wall}=S_{wall}(x,y)$.

Which terms in the equations of motion will contribute to the boundary stress-energy? When integrating the equations of motion across the wall, the first derivative of any function of $r$ will not contribute, whereas the second derivative will:
\bea
&&\int_{r_0-\epsilon}^{r_0+\epsilon}dr f'(r)=f(r_0+\epsilon)-f(r_0-\epsilon)=0;\\
&&\int_{r_0-\epsilon}^{r_0+\epsilon}dr f''(r)=f'(r_0+\epsilon)-f'(r_0-\epsilon)=-2f'(r_0-\epsilon).
\eea
Therefore, in order to solve for the action at the wall, we only need to consider the terms in the equations of motion which have second derivatives of functions of $r$. At zeroth order in $\epsilon$ the gauge field is independent of $r$, and the Einstein equations only depend on up to first derivatives of the metric functions. In this case, integrating the equations across the wall we find no contributions, and we find that we do not need an $S_{wall}$ at zeroth order in $\epsilon$.

At first order in $\epsilon$, we need to consider integrating the $\psi$ equation of motion across the wall. We will add the term
\be
\label{wallmass}
S_{wall}^\psi=\int\limits_{r=r_0} d^3x\sqrt{-h}~\delta m_w^2|\psi|^2
\ee
to the action, where $h_{\mu\nu}$ is the induced metric at $r=r_0$, and $\delta m_w$ is a localized shift in the mass of $\psi$. The nonzero contributions to the equation of motion when integrated over the wall are
\be
-\int_{r_0-\epsilon}^{r_0+\epsilon} dr \sqrt{-g}g^{rr}\psi''=\int_{r_0-\epsilon}^{r_0+\epsilon}  dr \sqrt{-h}~\delta m_w^2\psi\delta(r-r_0),
\ee
which gives the result $\delta m_w^2=2\alpha/L$.

Note that after adding the wall-localized mass term (\ref{wallmass}), the strategy for finding the critical field at which a phase transition occurs
is the following.  For a fixed choice of the wall localized mass and the location of the wall $r_0$, there is a critical value of the $B$-field at which
the purely normalizable solution for $\psi$ obeys the boundary conditions.  In this paper, we are always expanding about this critical field,
with $\epsilon$ parametrizing the distance from criticality.

We note also that we will need to add additional terms to $S_{wall}$ when we consider the equations of motion at $O(\epsilon^2)$ in the next section.

\subsection{Higher order corrections to the gauge field and metric}

The relevant equations of motion are the Einstein equations and the Euler-Lagrange equations for $\psi,A_\mu$, equations (\ref{eq:psi_eom}) and (\ref{eq:A_eom}).

In an $AdS_2\times R^2$ background, all the unknown functions scale as power laws in $r$. At $O(\epsilon)$ there is only the $\psi$ equation of motion. From \S3 we know there exists a lattice solution of the form
\be
\psi_1(r,x,y)=\rho_0r^\alpha\sum_{l=-\infty}^\infty e^{\frac{2\pi i l\sqrt{Q}x}{v_1}}e^{-{Q\over 2}\left (y+\frac{2\pi il}{v_1\sqrt{Q}}\right )^2},
\ee
where $\rho_0$ is the magnitude of $\psi_1$ and the scaling exponent is
\be
\alpha={1\over 2}\left(1+\sqrt{1+4(Q+m^2L^2)}\right).
\ee

$\psi_1$ acts as a source in the gauge field equation of motion and Einstein equations at $O(\epsilon^2)$. Therefore we can extract the $r$ scaling in the $O(\epsilon^2)$ corrections and solve the equations of motion for the spatial dependence. We write
\be
f_i(r,x,y)= \rho_0^2 r^{2\alpha}f_i(x,y)
\ee 
where $f_i=a,b,a_2^x,a_2^y$. By assuming a normalizable radial dependence of this form for each field, we are implicitly setting one integration constant to zero per function. Our choice of solution for each of these fields will also fix the form of the localized stress energy we will need to add at the wall in order to have a consistent solution.

Now we examine the differential equations at $O(\epsilon^2)$. The equation of motion for $A_r$ gives us the constraint
\be
\partial_xa_2^x+\partial_ya_2^y=0.
\ee
Besides this, we have 2 additional gauge field equations of motion (one each for $x,y$) and 5 nontrivial Einstein equations (for $G_{tt},G_{rr},G_{xx}=G_{yy},G_{rx},G_{ry}$) at $O(\epsilon^2)$. These are seven equations and four unknown functions. Luckily, three of them are redundant and we can find a consistent solution once we have chosen the form of the source, $\psi_1$.
The Einstein equations are
\bea\nonumber
2(\partial_ya_2^x-\partial_xa_2^y)+Q(\partial_x^2+\partial_y^2)(a+b)+2Q(4\alpha^2-1)b&=&S_1(\psi_1)\\\nonumber
2(\partial_xa_2^y-\partial_ya_2^x)+Q(\partial_x^2+\partial_y^2)(a-b)+2Q(2\alpha+1)b&=&S_2(\psi_1)\\\nonumber
2\alpha a_2^y+Q(\alpha-1)\partial_xa-Q\alpha\partial_xb&=&S_3(\psi_1)\\\nonumber
2\alpha a_2^x-Q(\alpha-1)\partial_ya+Q\alpha\partial_yb&=&S_4(\psi_1)\\
2(\partial_xa_2^y-\partial_ya_2^x)-2Q(\alpha-1)(2\alpha-1)a+2Q(2\alpha^2-\alpha+1)b&=&S_5(\psi_1)
\eea
and the gauge field equations are
\bea\nonumber
(\partial_x^2+\partial_y^2+2\alpha(2\alpha-1))a_2^x-Q\partial_yb&=&S_6(\psi_1)\\
(\partial_x^2+\partial_y^2+2\alpha(2\alpha-1))a_2^y+Q\partial_xb&=&S_7(\psi_1),
\eea
where the $\psi$-dependent source terms are given by
\bea\nonumber
S_1(\psi_1)&=&-{Q\over 2}(2\alpha^2+Q^2m^2+2Q^2y^2)|\psi_1|^2+iQ^2y(\psi_1^*\partial_x\psi_1-\psi_1\partial_x\psi_1^*)-Q(|\partial_x\psi_1|^2+|\partial_y\psi_1|^2)\\\nonumber
S_2(\psi_1)&=&{Q\over 2}(-2\alpha^2+Q^2m^2+2Q^2y^2)|\psi_1|^2-iQ^2y(\psi_1^*\partial_x\psi_1-\psi_1\partial_x\psi_1^*)+Q(|\partial_x\psi_1|^2+|\partial_y\psi_1|^2)\\\nonumber
S_3(\psi_1)&=&{Q\alpha\over 2}\partial_x|\psi_1|^2\\\nonumber
S_4(\psi_1)&=&-{Q\alpha\over 2}\partial_y|\psi_1|^2\\\nonumber
S_5(\psi_1)&=&-{Q\over 2}(2\alpha^2+Q^2m^2+2Q^2y^2)|\psi_1|^2+iQ^2y(\psi_1^*\partial_x\psi_1-\psi_1\partial_x\psi_1^*)-Q(|\partial_x\psi_1|^2-|\partial_y\psi_1|^2)\\\nonumber
S_6(\psi_1)&=&-{i\over2}Q^2(\psi_1^*\partial_x\psi_1-\psi_1\partial_x\psi_1^*+2iyQ|\psi_1|^2)\\
S_7(\psi_1)&=&{i\over2}Q^2(\psi_1\partial_y\psi_1^*-\psi_1^*\partial_y\psi_1)
\eea
and $a,b,a_2^x,a_2^y,\psi_1$ are now only functions of $x,y$ as we have omitted the power law $r$-dependence. 

We know that the vortex lattice solution is periodic in $x,y$ with periodicity ${v_1\over \sqrt{Q}}$ in the $x$ direction and ${2\pi\over v_1\sqrt{Q}}$ in the $y$ direction (this is only for the rectangular lattice); therefore we can expand each of these functions as a double Fourier series in $x,y$ as
\be\label{eq:xy_exp}
f_i(x,y)=\sum_{k,l} v_1e^{{2\pi ik\sqrt{Q}x\over v_1}}e^{ilv_1\sqrt{Q}y}e^{-\frac{k^2\pi^2}{v_1^2}-i\pi kl-\frac{l^2v_1^2}{4}}\tilde{f_i}(k,l),
\ee
where $f_i=a,b,a_2^x,a_2^y$, and we have pulled out the exponential function of $m,n$ which will be present in all of the source terms. Notice that the periodicity implies that each unit cell has a net flux density of $2\pi\over Q$.  It remains to Fourier transform the source terms in the equations of motion in order to bring them into the form of equation (\ref{eq:xy_exp}), and then solve algebraic equations for the polynomial coefficients $\tilde{f_i}(k,l)$. In order to do this we will use properties of exponentials and the Fourier transform to write an infinite sum of Gaussians as an infinite sum of exponentials,
\be
\sum_ke^{-{1\over 2}\left(y+{2\pi k\over v_1}\right )^2}e^{-{1\over 2}\left(y+{2\pi (k+l)\over v_1}\right )^2}=\sum_k e^{iv_1ky}{v_1\over 2\sqrt{\pi}}e^{-{v_1^2k^2\over 4}-i\pi kl-{l^2\pi^2\over v_1^2}}.
\ee

First we will do this for $Q=1$. In this case we also have $L=1/\sqrt{2}$, $\Lambda=-1$, and $m^2=2(\alpha^2-\alpha-1)$. Plugging in our ansatz of equation (\ref{eq:xy_exp}), we get the following algebraic equations for the $\tilde{f_i}(k,l)$:
\bea\label{eq:coeffs}\nonumber
2i\left (lv_1\tilde{a_2^x}-\frac{2\pi k}{v_1}\tilde{a_2^y}\right )+2(4\alpha^2-1)\tilde{b}-\left [\left ({2\pi k\over v_1}\right )^2+(lv_1)^2\right ](\tilde{a}+\tilde{b})&=&\frac{\alpha(1-2\alpha)}{2\sqrt{\pi}}+\frac{\pi^{3/2}k^2}{v_1^2}+\frac{l^2v_1^2}{4\sqrt{\pi}}\\\nonumber
2i\left (\frac{2\pi k}{v_1}\tilde{a_2^y}-lv_1\tilde{a_2^x}\right )+2(2\alpha+1)\tilde{b}-\left [\left ({2\pi k\over v_1}\right )^2+(lv_1)^2\right ](\tilde{a}-\tilde{b})&=&-\frac{\alpha}{2\sqrt{\pi}}-\frac{\pi^{3/2}k^2}{v_1^2}-\frac{l^2v_1^2}{4\sqrt{\pi}}\\\nonumber
2\alpha \tilde{a_2^y}+\frac{2\pi i k}{v_1}(\alpha(\tilde{a}-\tilde{b})-\tilde{a})&=&\frac{i\alpha\sqrt{\pi}k}{2v_1}\\\nonumber
2\alpha \tilde{a_2^x}+ilv_1(\tilde{a}-\alpha(\tilde{a}-\tilde{b}))&=&-\frac{il\alpha v_1}{4\sqrt{\pi}}\\\nonumber
2i\left (\frac{2\pi k}{v_1}\tilde{a_2^y}-lv_1\tilde{a_2^x}\right )-2(\alpha-1)(2\alpha-1)\tilde{a}+2(2\alpha^2-\alpha+1)\tilde{b}&=&\frac{1+\alpha-2\alpha^2}{2\sqrt{\pi}}\\\nonumber
\left [2\alpha(2\alpha-1)-\left ({2\pi k\over v_1}\right )^2-(lv_1)^2\right ]\tilde{a_2^x}-ilv_1\tilde{b}&=&-\frac{ilv_1}{4\sqrt{\pi}}\\
\left [2\alpha(2\alpha-1)-\left ({2\pi k\over v_1}\right )^2-(lv_1)^2\right ]\tilde{a_2^y}+{2\pi ik\over v_1}\tilde{b}&=&\frac{ik\sqrt{\pi}}{2v_1}.
\eea

From this we can see that we expect $\tilde{a}$ and $\tilde{b}$ to be real and $\tilde{a_2^x}$ and $\tilde{a_2^y}$ to be pure imaginary. For $k=l=0$ we get the solution
\be
\tilde{a}=\frac{\alpha(2\alpha^2+\alpha-4)-1}{4(\alpha-1)(4\alpha^2-1)\sqrt{\pi}},~~\tilde{b}=-\frac{\alpha}{4(2\alpha+1)\sqrt{\pi}},~~\tilde{a_2^x}=\tilde{a_2^y}=0,
\ee
and in all other cases the solutions are
\bea\nonumber
&&\tilde{a}(k,l)=-\frac{\alpha v_1^2((\alpha+1)4k^2\pi^2+v_1^2(2+(\alpha+1)l^2v_1^2-2\alpha(2\alpha^2+\alpha-4)))}{D}\\\nonumber
&&\tilde{b}(k,l)=-\frac{16k^4\pi^4+(8k^2\pi^2v_1^2+2l^2v_1^6)(1+l^2v_1^2+\alpha(2-3\alpha))+v_1^4(-l^4v_1^4+4\alpha^2(2\alpha^2-3\alpha +1))}{2D}\\\nonumber
&&\tilde{a_2^x}(k,l)=\frac{ilv_1^3(4k^2\pi^2+v_1^2(1+l^2v_1^2+(2-3\alpha)\alpha))}{D}\\
&&\tilde{a_2^y}(k,l)=-\frac{2\pi ikv_1(4k^2\pi^2+v_1^2(1+l^2v_1^2+(2-3\alpha)\alpha))}{D},
\eea
where 
\be
D=2\sqrt{\pi}(16k^4\pi^4+8k^2\pi^2v_1^2(l^2v_1^2+2\alpha(1-2\alpha))+v_1^4(l^4v_1^4+4l^2v_1^2\alpha(1-2\alpha)+4\alpha(\alpha-1)(4\alpha^2-1))).
\ee
Note that the equations of (\ref{eq:coeffs}) are not solvable for $\alpha=\pm {1\over 2}, 1$. In the case we have chosen, where $Q=1$, $\alpha =\frac{1}{2}(1+\sqrt{5+2m^2})$. Thus, we can only solve these equations for some values of $m$. 

\subsection{$O(\epsilon^2)$ stress-energy at the wall}

At $O(\epsilon^2)$ we need to consider the gauge field equations of motion and the Einstein equations integrated across the wall. We will now consider the following action at the wall:
\be
\label{wallack}
S_{wall}=\int\limits_{r=r_0} d^3x\sqrt{-h}\left \{\delta m_w^2|\psi|^2+ \epsilon^2A_\mu J_w^\mu + \epsilon^2(T_w)_\mu^{~\mu}\right \},
\ee
where we have added a current $J_w^\mu$ which couples to the gauge field, as well as a source of stress-energy $(T_w)_{\mu\nu}$ localized at the wall. This is the most general form of action we can add to the wall and should easily lead to a solution. Because we don't want the boundary current or stress tensor to enter into the equations of motion at zeroth order, we have assumed that each term enters the action at $O(\epsilon^2)$.

First we will consider integrating the gauge field equations of motion across the wall. The relevant equations are those for $A_x,A_y$. The equations we must solve are
\be
\int_{r_0-\epsilon}^{r_0+\epsilon} dr \sqrt{-g}g^{rr}g^{xx}(a_2^{r,x,y})''=\int_{r_0-\epsilon}^{r_0+\epsilon}  dr \sqrt{-h}h^{xx} (J_w)_{x,y}\delta(r-r_0),
\ee
which have the solutions
\be
 (J_w)_{x,y}=-\frac{4\alpha}{L}a_2^{x,y}(r_0,x,y).
\ee

Finally, we must consider the Einstein equations. There are three equations which include second derivatives of the fields, $G_{tt}, G_{xx},$ and $G_{yy}$. The total stress-energy from the action at the wall takes the form
\be
-{\sqrt{-h}h_{\mu\nu}\over 2}\mathcal L_{wall}+\epsilon^2\sqrt{-h}(A_\mu (J_w)_\nu+(T_w)_{\mu\nu})
\ee
where $\mathcal L_w$ is the integrand of $S_{wall}$. After integrating the Einstein equations, we get the following set of equations for $T_w$:
\bea\nonumber
&&{\alpha\over r_0^3}b(r_0,x,y)={L^5\over 2r_0^3}\left (\delta m_w^2|\psi_1(r_0,x,y)|^2+{Qy\over L^2}(J_w)_x+{r_0^2\over L^2}(T_w)_{tt}+{1\over L^2}(T_w)_{xx}+{1\over L^2}(T_w)_{yy}\right )\\\nonumber
&&{\alpha\over 2 r_0}(a(r_0,x,y)-b(r_0,x,y))=-{L^5\over 2r_0}\left (\delta m_w^2|\psi_1|^2-{Qy\over L^2}(J_w)_x-{r_0^2\over L^2}(T_w)_{tt}-{1\over L^2}(T_w)_{xx}+{1\over L^2}(T_w)_{yy}  \right )\\\nonumber
&&{\alpha\over 2 r_0}(a(r_0,x,y)-b(r_0,x,y))=-{L^5\over 2r_0}\left (\delta m_w^2|\psi_1|^2+{Qy\over L^2}(J_w)_x-{r_0^2\over L^2}(T_w)_{tt}+{1\over L^2}(T_w)_{xx}-{1\over L^2}(T_w)_{yy}  \right ).
\eea
which have the solution
\bea\nonumber
&&(T_w)_{tt}={\alpha\over r_0^2L^3}(a(r_0,x,y)-b(r_0,x,y))+{2\alpha L\over r_0^2}|\psi_1(r_0,x,y)|^2\\\nonumber
&&(T_w)_{xx}=(T_w)_{yy}+{4\alpha Qy\over L}a_2^x(r_0,x,y)\\
&&(T_w)_{yy}=-2\alpha L|\psi_1(r_0,x,y)|^2+{\alpha\over 2L^3}(3b(r_0,x,y)-a(r_0,x,y)).
\eea

We note that, as with the original Randall-Sundrum matching \cite{RS}, the wall-localized stress-energy violates the Null Energy Condition.  This is
not a significant concern here (as it was not there); warped solutions microscopically realizing Randall-Sundrum like warping have been found in the full string
theory, and we expect similar solutions could be found in this more involved case.  It does mean that the wall should not be considered as a 
`brane' which has Goldstone modes that allow it to fluctuate in the transverse dimensions.

\subsection{Pictures of the modulated phase}

We conclude this section with representative plots of the scalar supporting the vortex lattice $\psi_1(x,y)$ (Figure 1), the
modulation of flux density in the crystal (Figure 2), and a representative crystalline metric function (plotted as a function of (x,y) in Figure 3
and (r,y) in Figure 4).  All plots are for values of the parameters given by: $Q=1$, $\alpha = {1\over 2} + {{\sqrt 3} \over 2}$, $v_1 = \sqrt{2\pi}$ (a
square lattice).  The functions have been approximated keeping 121 terms in the Fourier series (i.e., with $k,l$ running from $-5$ to $5$ in the
formulae above).

\begin{figure}
\begin{center}
\includegraphics{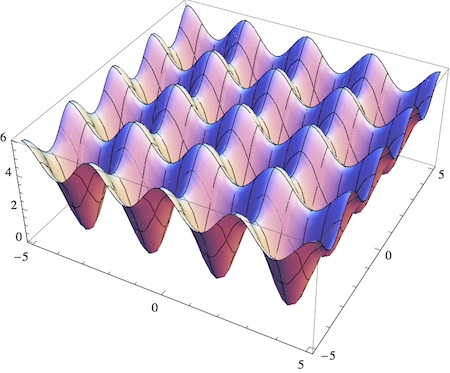}
\caption{The scalar vortex lattice configuration $\psi_{1}(x,y)$.}
\end{center}
\end{figure}

\begin{figure}
\begin{center}
\includegraphics{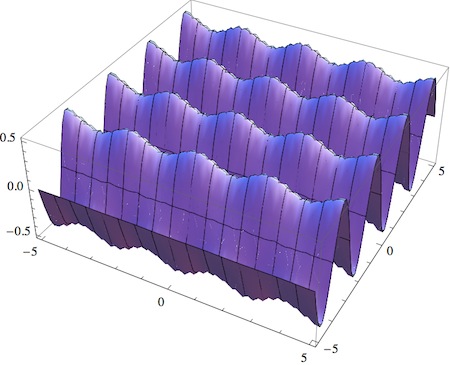}
\caption{The ${\cal O}(\epsilon^2)$ correction to the $A_x$ gauge field, which controls the modulation of the flux density in the lattice.}
\end{center}
\end{figure}

\begin{figure}
\begin{center}
\includegraphics{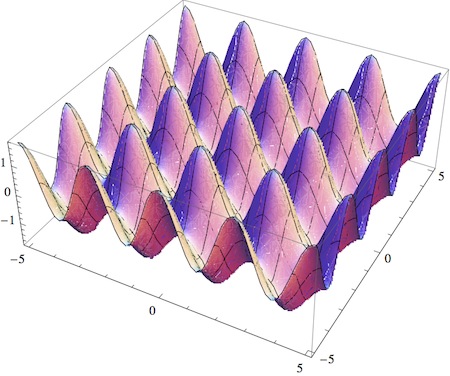}
\caption{The metric function $b(x,y)$ generated by the back-reaction of a square vortex lattice.}
\end{center}
\end{figure}

\begin{figure}
\begin{center}
\includegraphics{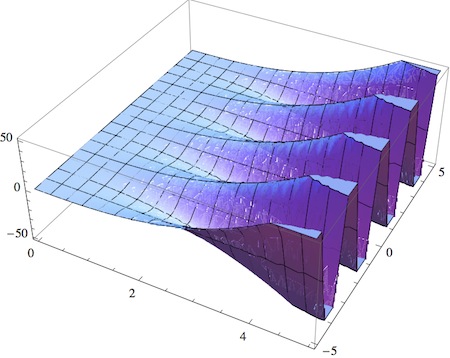}
\caption{The metric function in front of $dx^2 + dy^2$, now plotted as a function of $y$ and $r$.  We have chosen $x=2$ for this plot.}
\end{center}
\end{figure}

\section{Comments on physics of the lattice model}

From the form of the deformed metric in \S3, we can infer some basic facts about the physics of the lattice solution. The IR wall geometry
we have implemented is a bottom-up implementation of IR confinement \cite{PS}.  In physical observables, powers of the IR radial cutoff
$r_0$ can be replaced by powers of $1/\Lambda$, with $\Lambda$ the scale of confinement.  However, it is common in such solutions that
also at finite temperature, one could (after the transition from confinement to deconfinement represented by a horizon at some $r < r_0$) 
replace powers of $r_0$ by $1/T$.  Using this correspondence, we can infer the leading corrections to thermodynamic functions.

The free energy density ${\cal F}$ will receive a correction at ${\cal O}(\epsilon^2)$.  It will have the general form
\begin{equation}
\label{basicscaling}
{\cal F} \sim T \left(1 + \epsilon^2 T^{-2\alpha} + \cdots \right)
\end{equation}
where the leading term comes from the $AdS_2$ geometry (and gives rise to the notorious extensive ground-state entropy), and the
subleading term is due to the physics of the vortex lattice.  
One can see that the ${\cal O}(\epsilon^2)$ corrections will scale like $\Lambda^{-2\alpha}$ in the confining geometry
quite explicitly, both from the form of the wall action (\ref{wallack}), and from the $\epsilon$ expansion
of the contributions to the bulk action.

The schematic formula (\ref{basicscaling}) makes it clear that for a given value
of $\epsilon$ and $\alpha$, there is an IR scale beneath which one should not trust perturbation theory.  To avoid this region, one must keep
\begin{equation}
T > \epsilon^{1\over \alpha}~.
\end{equation}
As $\alpha$ increases, the regime of trustworthiness of the linearized solution shrinks; this is in keeping with the simple intuition that the
perturbation expansion in powers of $\epsilon r^{\alpha}$ will break down at smaller values of $r$ for larger $\alpha$.

Free classical defects would contribute a correction to the free energy density proportional to $T$ (and, of course, inversely proportional to
the lattice spacing).  The exponent $\alpha$ therefore parametrizes an anomalous scaling of the free energy per vortex, characteristic of
the strongly coupled field theory.

What happens beyond the regime where perturbation theory around the transition is valid?  One natural speculation is that as one proceeds
to the deep IR, the different lattice sites `decouple' in a manner similar to that seen in $AdS_2$ fragmentation \cite{fragment}.  
It is possible that this would proceed via another phase transition (at a temperature/energy scale lower than
the transition to the vortex lattice state) to a `fragmented' state.
Such a fate was proposed in \cite{Subir} for the D-brane lattice models of \cite{branelattices}, where it was speculated that this might also
characterize the physics of generic $AdS_2$ horizons.  The growing localization of the dominant contribution to the low-temperature
entropy on distinct lattice sites in the gravity solution provides support for this idea, in perturbation theory. 

Finally, it is worth emphasizing that the lattices discussed here are quite distinct from those obtained in related literature by considering
a periodic spatial variation in the chemical potential $\mu(x)$ \cite{Horowitz}.
The key difference lies in the nature of the IR behavior.  In systems with a finite charge density,
spatial modulations of $\mu$ can and will be cancelled by the background charge carriers -- they will be screened.  The hard lattices of
the sort discussed here, in contrast, cannot be screened (physically, one cannot screen a magnetic field), and their effects should be expected
to persist to the deep IR. In the S-dual perspective, 
such a feature is natural for the analog of `Wigner crystallization' of charged carries that are added to 
a conformal field theory.

\section{Connecting with more general gravity solutions}

Here, we describe how the lattice solutions we found in \S3\ should also arise in `IR completions' of metrics with rather general dynamical critical exponent $z$ and hyperscaling violation parameter $\theta$ \cite{GKPT,Kiritsis,tadashi1,hyper}.  
The basic point will be that, as in \cite{HKW} and \cite{Cremonini2}, the $AdS_2$
can arise in the deep IR, where corrections to the action supporting such solutions can become important.

The bigger physics picture is the following.  As discussed in \cite{Sachdevnew}, one can expect expectation values of monopole operators to serve as order parameters for translation-breaking phases in doped critical field theories.  By S-duality in the 4d bulk, one can map the magnetically charged field dual to the monopole operator, to an electrically charged field.  The doping maps to a background magnetic field.  Then, the lattices found in \S3\ give concrete examples of the solids described in \cite{Sachdevnew}, in strongly coupled quantum field theory.  The considerations of this section show that this can happen in models with rather general $z$ and $\theta$.

\subsection{Basic EMD theory and magnetic solutions}

We start with the bulk gravity theory represented by an Einstein-Maxwell-Dilaton action
\begin{equation}
\label{bulkac}
S = \int d^4x \sqrt{-g} \left( R - 2 (\partial \phi)^2 - f(\phi) F_{\mu\nu}F^{\mu\nu} - V(\phi) \right)
\end{equation}
where the gauge-coupling function is of the form
\begin{equation}
f(\phi) = e^{2\alpha\phi}
\end{equation}
and the scalar potential takes the form
\begin{equation}
V(\phi) = {1\over L^2} e^{-\eta \phi}~.
\end{equation}
This theory supports solutions of the form
\begin{equation}
ds^2 = L^2 \left(-\overline{a}(r)^2 dt^2 + {dr^2 \over \overline{a} (r)^2} + \overline{b} (r)^2 (dx^2 + dy^2)\right)
\end{equation}
with scalar profile
\begin{equation}
\phi(r) = K ~{\rm log}(r)~.
\end{equation}

In the simplest solutions, $\overline{a}$ and $\overline{b}$ take power-law scaling forms, and one
finds
\begin{equation}
\label{abscaling}
{\overline a(r)} = C_a~ r^{1-\gamma},~\overline{b}(r) = C_b~ r^{\beta}~
\end{equation}
in the vicinity of the horizon (with differences arising as one goes towards the UV, if one wishes to find asymptotically
AdS solutions, as in e.g. \cite{GKPT}; note that we use the convention that $r \to \infty$ is the UV  and $r \to 0$ is the IR in this section, in contrast with earlier sections).  
The two exponents $\gamma$ and $\beta$, fixed by the parameters $\alpha$ and $\eta$ in the action,  capture the scaling properties of the IR fixed point induced by doping the CFT.
The physically relevant dynamical critical exponent $z$ and the hyperscaling violation exponent $\theta$ are determined in terms of $\gamma$ and $\beta$
by the equations
\begin{equation}
\beta = {{\theta -2} \over {2(\theta-z)}},~\gamma = {\theta \over {2(\theta - z)}}~.
\end{equation}
General values of these exponents were first obtained in dilatonic systems in \cite{Kiritsis}.

In many cases that have been studied, these solutions can in fact be supported in two different ways.  If one studies an electrically charged black brane,
Gauss' law yields the solution
\begin{equation}
\label{efield}
F = {Q_{e} \over f(\phi) {\overline b(r)}^2} dt \wedge dr~.
\end{equation}
One then finds extremal solutions where $K < 0$ and $\phi \to \infty$ near the horizon $r \sim 0$.   This means that
the coupling is vanishing.  As discussed in \cite{GKPT}, in the very near-horizon regime, the solution is then unreliable;
in a full UV complete theory like string theory, higher derivative corrections will usually become important, because new light states
appear as $g = e^{\alpha\phi} \to 0$.   This difficulty can be avoided by turning on a small temperature, since this cuts off the
running of the dilaton; and the near-horizon solutions for finite $T$ are simple to write down as well.

However, in a 4d bulk, one can also use bulk electric-magnetic duality to find a representation of the solution in terms of a 
magnetically charged black brane, i.e. a field theory immersed in a background magnetic field.  This allows us to make contact
with our discussion in \S2 and \S3, and with the picture of \cite{Sachdevnew}:

\medskip
\noindent
$\bullet$ Suppose one is interested in studying the physics of monopole operators to diagnose the phase structure of the `electric' model.
One could introduce monopoles into the theory (\ref{bulkac}) and compute their correlators
using semi-classical techniques in a multi-soliton background.  However, it is easier to realize that by electric/magnetic duality,
one can represent the monopoles as quanta of fundamental electrically charged fields in a dual theory, where the electric background
(\ref{efield}) is dualized to a background magnetic field. 

\medskip
\noindent
$\bullet$ As mentioned above, the running dilaton indicates an `IR incompleteness' of the solution -- as the dilaton runs to extreme
values, new corrections typically become important and deform the solution.  For magnetically charged black branes in these dilatonic system,
one possible result of the corrections is the emergence, in the deep IR, of an $AdS_2$ geometry. This was discussed for Lifshitz scaling metrics
in \cite{HKW} and for general $\theta$ and $z$ in \cite{Cremonini2}.

The end result is that in critical theories dual to dilatonic systems with fairly generic $z$ and $\theta$, if we are concerned mostly with the 
physics at very low energies, we can study monopole operators by
considering the dynamics of electrically charged scalars in an $AdS_2$ throat supported by magnetic flux.  This provides a rather general
setting where our analysis in \S3\ could be of relevance.

\section{Discussion}

There are many interesting directions for future exploration of the analytical vortex lattice solutions described here.  We briefly mention some of these now.

\medskip
\noindent
$\bullet$ It should be possible to find analogous perturbative crystalline
geometries emerging directly out of solutions with various values of $z$ and $\theta$, without invoking the transition to an $AdS_2$ space-time
\cite{WIP}. 

\medskip
\noindent
$\bullet$ It would be natural to explore replacing the IR Israel thin wall considered here, with a black brane horizon.

\medskip
\noindent
$\bullet$ One would like to compute simple correlation functions in these backgrounds.  For instance, quasi-universal features have been seen
in the transport properties of simple holographic lattice models in \cite{Horowitz}.  Their analogues in this system are worth exploring \cite{WIP}.

\medskip
\noindent
$\bullet$ Most ambitiously, it would be nice to find the full non-linear solution to the coupled set of partial differential equations that
characterize the system.  This would most likely rely on powerful numerical techniques. This program should yield new insights
on the `fragmentation' phenomenon, and the eventual emergence of a solid in a `confined' phase of the boundary gauge theory.

\bigskip
\centerline{\bf{Acknowledgements}}

We would like to thank G. Horowitz and A. Karch for helpful discussions. SK also thanks S.~Yaida for enjoyable discussions about crystalline horizons in the
summer of 2010.   This project was initiated when SK and SS were hosted by the Simons Foundation in New York.  SK and SS would also like to acknowledge the hospitality
of the Simons Symposium on `Quantum Entanglement: From Quantum Matter to String Theory', 
and thank the participants for stimulating discussions.  SK is supported by the U.S.\ National Science Foundation grant
PHY-0756174, the Department of Energy under contract  DE-AC02-76SF00515, and the John Templeton Foundation.
SS is supported by NSF under grant DMR-1103860, by the U.S.\ Army Research Office Award W911NF-12-1-0227,
and by the John Templeton Foundation.

\end{document}